\documentclass[aps,prl,twocolumn,groupedaddress,preprintnumbers,nofootinbib]{revtex4}

\usepackage[dvips]{graphicx}
\usepackage{color}
\usepackage{amsmath,amssymb,slashed}
\newcommand{\eqqa}{\begin{equation}\begin{aligned}}
\newcommand{\eqqae}{\end{aligned}\end{equation}}
\newcommand{\beqs}{\begin{equation*}}
\def\beq{\begin{equation}}
\newcommand{\la}{\lambda}

\newcommand{\braket}[2]{\langle #1  #2 \rangle}

\newcommand{\eeqs}{\end{equation*}}
\def\eeq{\end{equation}}

\newcommand{\beqas}{\begin{eqnarray*}}
\newcommand{\beqa}{\begin{eqnarray}}

\newcommand{\eeqas}{\end{eqnarray*}}
\newcommand{\eeqa}{\end{eqnarray}}

\newif\ifdraft
\drafttrue
\newif\ifpreprint
\preprinttrue
\def\spa#1.#2{\left\langle#1\,#2\right\rangle}
\def\spb#1.#2{\left[#1\,#2\right]}

\newcommand{\eq}{\begin{equation}}
\newcommand{\eqe}{\end{equation}}
\newcommand{\eqa}{\begin{eqnarray}}
\newcommand{\eqae}{\end{eqnarray}}

\newcommand{\bea}{\begin{eqnarray}}
\newcommand{\eea}{\end{eqnarray}}

\newbox\charbox
\newbox\slabox
\def\s#1{{      % Feynman slash
        \setbox\charbox=\hbox{$#1$}
        \setbox\slabox=\hbox{$/$}
        \dimen\charbox=\ht\slabox
        \advance\dimen\charbox by -\dp\slabox
        \advance\dimen\charbox by -\ht\charbox
        \advance\dimen\charbox by \dp\charbox
        \divide\dimen\charbox by 2
        \raise-\dimen\charbox\hbox to \wd\charbox{\hss/\hss}
        \llap{$#1$}
}}

\begin{document}
\preprint{KOBE-COSMO-19-03, MAD-TH-19-01}
\preprint{QMUL-PH-19-03,  NCTS-TH/1901}

\title{Unitarity bounds on charged/neutral state mass ratio} 

\author{
Wei-Ming Chen,$^{1}$  Yu-tin Huang,$^{2,3}$ Toshifumi Noumi,$^{4,5}$  Congkao Wen,$^{6}$}
\affiliation{$^1$ Department of Physics , National Tsing-Hua University,
No.101, Section 2, Kuang-Fu Road, Hsinchu, Taiwan}
\affiliation{$^2$ Department of Physics and Astronomy, National Taiwan University, Taipei 10617, Taiwan}
\affiliation{$^3$ Physics Division, National Center for Theoretical Sciences, National Tsing-Hua University,
No.101, Section 2, Kuang-Fu Road, Hsinchu, Taiwan}
\affiliation{$^4$ Department of Physics, Kobe University, Kobe 657-8501, Japan}
\affiliation{$^5$ Department of Physics, University of Wisconsin-Madison, Madison, WI 53706, USA}
\affiliation{$^6$ Centre for Research in String Theory, School of Physics and Astronomy,
Queen Mary University of London, Mile End Road, London E1 4NS, United Kingdom}

\begin{abstract}
In this letter, we study the implications of unitary completion of quantum gravity on the low energy spectrums, through an infinite set of unitarity bounds on the forward-limit scattering amplitudes. In three dimensions, we find that light states with charge-to-mass ratio $z$ greater than $1$ can only be consistent if there exists other light states, preferably neutral. Applied to the compactification of the Standard Model, where the low energy couplings are dominated by the electron with $|z|\sim 10^{22}$, this provides a novel understanding of the need for light neutrinos. 
\end{abstract}

\maketitle

%%%%%%%%%%%%%%%%%%%%%%%%%%%%%%%%%%%%%%%%%%%%%%
\section{Introduction}  
%%%%%%%%%%%%%%%%%%%%%%%%%%%%%%%%%%%%%%%%%%%%%%
Quantum consistency of scattering amplitudes such as unitarity and analyticity provides stringent constraints on the particle contents and interactions of the theory. For example, unitarity requires that amplitudes should not blow up at high energy, which has been used as a consistency condition to predict new particles at high energy. Together with analyticity, it also implies nontrivial conditions on the low-energy effective field theory (EFT). It has been known that Wilson coefficients have to respect certain positivity bounds~\cite{Adams:2006sv}. More recently, a new infinite set of positivity bounds on these coefficients was discovered in~\cite{NimaEFT}. In this letter we explore implications of the new positivity bounds on the charged (and neutral) state spectrum in quantum gravity, motivated by the weak gravity conjecture (WGC)~\cite{ArkaniHamed:2006dz}.

The WGC states that quantum gravity has to contain a charged state with a dimensionless charge-to-mass ratio $z$ bigger than unity~\cite{ArkaniHamed:2006dz}.
It has a wide range of phenomenological applications from cosmology to particle physics, hence it has been studied toward its proof from various perspectives, e.g., based on signs of Wilson coefficients~\cite{Kats:2006xp,Cremonini:2009ih,Cheung:2014ega,Andriolo:2018lvp,Cheung:2018cwt,Hamada:2018dde}. In particular, \cite{Cheung:2014ega} explored a possible connection between the WGC and the positivity bounds of~\cite{Adams:2006sv} on four-photon amplitudes by studying one-loop corrections from charged particles. Under several assumptions, it was argued that a bound on the charge-to-mass ratio~\cite{commentzstar},
\begin{align} \label{eq:cliff}
|z|>z_\ast\,,
\end{align}
similar to the WGC bound is obtained from positivity of the leading higher derivative correction to the Einstein-Maxwell theory. 

It is then natural to ask what are the implications of the stronger positivity bounds of~\cite{NimaEFT} to the charge-to-mass ratio. As we explain shortly, the unitarity in the UV implies a positive determinant of the $(\frac{n+2}{4}) \times (\frac{n+2}{4})$ Hankel matrix, constructed from the low-energy coefficients up to $2n$ derivatives ($n=2$  corresponds to the ordinary positivity bound used in~\cite{Cheung:2014ega}). Interestingly, in three dimensions, for an isolated light charge state this implies a bound of the form,
\begin{align}\label{eq:ab-bounds}
0\leq|z|<a\,,
\quad
b<|z|\,,
\end{align}
where the values of $a,b$ depend on the size of the matrix, and $z=\frac{qg \sqrt{M_{pl}}}{m}$ is the dimensionless charge-to-mass ratio in $d=3$. We find that $b$ linearly grows up as we increase the matrix size, whereas $a$ approaches to some value close to unity. It suggests that the bound is reduced to $|z|<\mathcal{O}(1)$ in the large-$n$ limit, which is surprisingly opposite to the WGC bound $|z|>\mathcal{O}(1)$. We interpret this observation as follows:
\begin{itemize}
\item In quantum gravity, a charged particle cannot have a charge-to-mass ratio $|z|>\mathcal{O}(1)$ without accompanied by other particles with $|z|<\mathcal{O}(1)$. 
\end{itemize}
Note that $|z|<\mathcal{O}(1)$ essentially means neutral as long as the gauge coupling is much bigger than the gravitational one. By further proceeding this analysis, we provide an infinite set of lower-bounds on the charged/neutral state mass ratio in three dimensions. It implies that if there exists a charged particle with $|z|>\mathcal{O}(1)$,
then there is a need for other light states with an upper bound on the mass.

%%%%%%%%%%%%%%%%%%%%%%%%%%%%%%%%%%%%%%%%%%%%%%
\section{The WGC from weakly interacting matter}  
%%%%%%%%%%%%%%%%%%%%%%%%%%%%%%%%%%%%%%%%%%%%%%
To motivate the plausibility of the constraints discussed above, let's consider the four-photon amplitude generated solely by weakly interacting states, with at most minimal coupling to the photon. In this case, the amplitudes are dominated by the one-loop effect, which we construct utilizing unitarity cuts~\cite{Bern:1994zx}, building loop integrands from tree amplitudes. Note that while gravitons have no on-shell degrees of freedom in three dimensions, the tree-level two-photon-two-matter amplitude due to graviton exchange is non-zero. This does not contradict with the usual statement that gravity does not produce long range force in three dimensions, since the residue of this amplitude vanishes much like that in Chern-Simons matter theory. The results of the one-loop amplitudes are expressed as, 
\begin{align}\label{OneLoopAmp}
\!\!\!\!\!\! M_4(s,t)& =4\, \mathcal{C}(34; 1, 2)   I_{\rm tri}(s,m^2)   
+  \mathcal{C}(12;34) I_{\rm bub}(s,m^2)  \nonumber
\\ & ~~~ + (s \leftrightarrow t)+ (s \leftrightarrow u) ,
\end{align}
where $I_{\rm bub}(s,m^2), I_{\rm tri}(s,m^2)$ are the scalar bubble and triangle integrals (with massive propagators) which constitute the integral basis in three dimensions. The explicit expressions of integral coefficients $\mathcal{C}(34; 1, 2), \mathcal{C}(12;34)$ for amplitudes of Einstein-Maxwell theory coupled with matters are given in the appendix. Here we consider the cases where the massive matter in the loop is a scalar or a fermion. As $s\rightarrow\infty$, in the forward limit we find that the amplitudes behave as,
\eqa
{\rm scalar}:\quad &&\frac{1-z^2}{7680 \pi m M^2_{pl}}s^2+\mathcal{O}(s^{\frac{3}{2}})\,,
\\ {\rm fermion}: \quad &&\frac{1-z^2}{2560 \pi m M^2_{pl}}s^2+\mathcal{O}(s^{\frac{3}{2}})\,.
\eqae
Note that the QED contribution, which would be proportional to $z^4$, is subdominant to the gravitational effects which behave as $s^2$ as $s\rightarrow \infty$.

We assume gravity is UV completed while weakly coupled. This means that we require the amplitude enjoy improved high energy behaviour, at least $<s^2$, order by order in $M_{pl}$. A canonical example would be perturbative string theory. Since the coefficient for $s^2$ is always positive for $|z|<1$ and negative for $|z|>1$, we immediately see that the spectrum must contain states that on both sides of the $|z|=1$ border, in particular $|z|>1$. This is precisely the content of the WGC.

Next, consider a spectrum that contains a light charged particle, with $|z| \gg 1$. This leads to a large negative contribution to the coefficient of $s^2$, which then requires states with even lighter masses and $|z|<1$ to compensate, preferably a neutral state. If the spectrum arrises from the compactification of our four-dimensional Standard Model, where the electron $|z|\sim 10^{22}$, the role of these light neutral states are taken up by the neutrinos~\cite{compactify}! Thus in this tentative example, we see that the UV completion of quantum gravity leads to correlation between charged states and light neutral states. In the next sections we will prove this connection in a more general setup.

The same analysis can also be applied to the case with multi-$U(1)$  gauge theories coupled with gravity. In this case, the photons can carry indices of different $U(1)$'s. It is convenient to consider a crossing-symmetric combination of amplitudes by multiplying them with auxiliary real unit vectors $u, v$~\cite{Andriolo:2018lvp}
\bea \label{eq:multi-U(1)}
M_4(s,t; u, v)= \sum_{i,j,k,l} u_i v_j v_k u_l M_4(1_i, 2_j, 3_k, 4_l),
\eea
where subscripts $i, j, k, l$ are the $U(1)$ indices. We now find that in the large-$s$ limit, the one-loop amplitudes behave as
\begin{align} \label{eq:multi-U1}
{\rm scalar}:\quad &&\frac{2-|\vec{z}\cdot u|^2-|\vec{z}\cdot v|^2}{7680 \pi m M^2_{pl}}s^2+\mathcal{O}(s^{\frac{3}{2}}), \\ 
{\rm fermion}: \quad && \frac{2-|\vec{z}\cdot u|^2-|\vec{z}\cdot v|^2}{2560 \pi m M^2_{pl}}s^2+\mathcal{O}(s^{\frac{3}{2}}) \,.
\end{align}
The same arguments of the single-$U(1)$ case (under the same assumptions) now imply that there must exist some state which satisfies
\bea
|\vec{z}\cdot u|^2 + |\vec{z}\cdot v|^2 > 2\,. 
\eea
Choose the unit vectors to be equal, $u=v$, the above condition implies the three-dimensional version of the convex-hull constraints $|\vec{z}\cdot u|^2  > 1$~\cite{Cheung:2014vva}. 

\vspace{-0.2cm}
%%%%%%%%%%%%%%%%%%%%%%%%%%%%%%%%%%%%%%%%%%%%%%
\section{Unitarity bounds on the forward amplitude}  
%%%%%%%%%%%%%%%%%%%%%%%%%%%%%%%%%%%%%%%%%%%%%%
As we discussed, we will be interested in the four-point amplitudes in the forward limit, namely $M_4(s,\theta)$ with $\theta=0$~\cite{forward-limit}. The amplitudes are analytic functions on the complex plane except for poles and branch cuts on the real $s$-axes, reflecting single or multi-particle productions.  Again we assume gravity is weakly coupled when it is UV completed. In practice, what this means is that we can sensibly talk about the amplitudes as perturbative series in $M_{pl}$, the three-dimensional Plank constant. In string theory, this would correspond to the case where $M_{string}\ll M_{pl}$. The non-dynamic nature of three-dimensional gravity is reflected in that the low energy effective theory is given by a theory of photons, with higher-dimensional operators generated by virtual gravitons and integrating away massive states. 

This motivates us to parametrize the low energy forward photon amplitudes at $\mathcal{O}(M_{pl}^{-2})$ as
\eq\label{3Dpara}
\left. M_4(s,0)\right|_{s/m_i^2 \ll1}=c_0\frac{s^{\frac{3}{2}}}{M^2_{pl}}{+}\sum_{i,n}\frac{\left(c_{n,4}z_i^4{+}c_{n,2}z_i^2{+}c_{n,0}\right)}{m_i^{2n{-}3} M^2_{pl}}\; s^{n}\,,
\eqe
where $n=2,4,6,\cdots$, and $i$ labels the massive states that were integrated out to obtain the EFT. The explicit coefficients can be computed from the one-loop amplitudes given in eq.(\ref{OneLoopAmp}) by expanding the massive loop integrals in the large mass limit, see e.g.~\cite{Chen:2015hpa}. For scalars we obtain: 
\begin{align}\label{ScalarCoeff}
 c_{n,4} &= { (n^2+n+1) \over 2^{2n + 5}  (n+ 2)(n+ 1) (2n+1)  }\, , \cr
c_{n,2} &=\begin{cases}
                {1 \over 7680}  \quad  \quad \quad \quad \quad \quad  \quad ~   {\rm if}  \quad  n=2 \\
 {(n + 1) \over 2^{2n+8} (2n+ 3) (2n+ 5) }  \quad ~~ \,  {\rm if}  \quad n>2
            \end{cases}\,,  \\
c_{n,0} &=\begin{cases}
                {1 \over 2560}  \quad  \quad \quad \quad \quad \quad \quad\quad \quad~\,   {\rm if}  \quad  n=2 \\
  { (n/2 + 1) (n + 1) \over 2^{2n+8} (2n+1) (2n+ 3) (2n+ 5) }  \quad  ~~ \!  {\rm if}  \quad n>2
            \end{cases}\,.  \nonumber 
\end{align}
 As for the fermions, we have: 
 \begin{align}\label{FermiCoeff}
 c_{n,4} &= { (5n^2+5n+2) \over 2^{2n + 5} n (n+2) (n+ 1) (2n+ 1) } \,, \cr
c_{n,2} &=\begin{cases}
                -{1 \over 3840}  \quad  \quad \quad \quad \quad \quad  \quad  \quad  \, \,  {\rm if}  \quad  n=2 \\
 {(n + 1) \over 2^{2n+8} (n+2)(2n+ 3) (2n+ 5) }  \quad \, \ {\rm if}  \quad n>2
            \end{cases} \,, \\
c_{n,0} &=\begin{cases}
                {1 \over 1920}  \quad  \quad \quad \quad \quad \quad \quad\quad \quad~ ~  {\rm if}  \quad  n=2 \\
{  (n + 1) \over 2^{2n+9}(2n+1) (2n+ 3) (2n+ 5) }   ~~~~\   {\rm if}  \quad n>2 \nonumber
            \end{cases}  \,.
\end{align}
The results for $n=2$ matches with those computed using the heat kernel method in~\cite{Ritz:1995nt, Drummond:1979pp, Andriolo:2018lvp}.
\vspace{0.2cm}

The leading term in eq.(\ref{3Dpara}) contains the massless cut from internal massless states such as photons. Here, we will simply work with the subtracted amplitude $\widetilde{M}_4=M_4-c_0\frac{s^{\frac{3}{2}}}{M^2_{pl}}$ such that the low energy coefficients are well defined:
\eq
\left. \widetilde{M}_4(s,0)\right|_{s/m_i^2 \ll1}=\sum_{n\in even}\; g_n\,s^n\,.
\eqe 

As pointed out in~\cite{Adams:2006sv}, the low-energy coefficients are related to the UV physics through  the analyticity of the S-matrix. This is exploited by considering the following integral:
\eq\label{Master}
I_n=\int_{\infty} \frac{ds}{s^{n+1}}\widetilde{M}_4\,,
\eqe
where the integration contour encircles the infinity. We assume that $\widetilde{M}_4$ is bounded as $<s^2$ in high energy, which is the case, e.g., when \eqref{3Dpara} is bounded by the Froissart bound $|M(s,0)|\leq s \log s$~\cite{Froissart:1961ux},~\cite{tpole}.
This implies that for a unitary completed theory,  $I_n=0$ for $n\geq 2$. Then eq.(\ref{Master}) leads to the conclusion that the contribution to the integral from the origin and those from the poles and discontinuities must cancel. In other words we have
\eq\label{Relate}
g_n=\sum_{a}\frac{p_a}{m^{2n+2}_a}+\sum_{b}\int_{4m_b^2} \frac{ds}{s^{n+1}}\, {\rm Im}\! \left[\widetilde{M}_4\right]\,,
\eqe
where $a,b$ labels all possible UV states that enter via tree-level exchange or loops respectively. Importantly $p_a$, which is the square of three-point couplings, and 
${\rm Im}\! \left[\widetilde{M}_4\right]$ are both positive. The former being the square of the three-point coupling, while the latter is the imaginary part of the forward amplitude, which is proportional to the cross section. 

Such constraints were discussed in~\cite{NimaEFT}, and there is a geometric interpretation: the coefficients $g_n$ must reside in the convex hull points on a half moment curve, i.e.,
\eq\label{Hull}
\left(\begin{array}{c} g_2 \\g_4 \\ g_6 \\ \vdots \end{array}\right)=\sum_{i} a_i \left(\begin{array}{c} x_i \\ x_i^2 \\ x_i^3 \\ \vdots \end{array}\right), \quad a_i \geq 0\,,
\eqe 
where $x_i\in R^+$. Organize the coefficients into a symmetric Hankel matrix:
\eq\label{Hankel}
K_{n}=\left(\begin{array}{cccc}g_2 & g_4 & g_6 & \cdots  \\ g_4 & g_6 & g_8 & \cdots  \\ g_6 & g_8 & g_{10} & \cdots \\ \vdots & \vdots & \vdots & \vdots \\ \cdots & \cdots & \cdots & g_n\end{array}\right)\,,
\eqe  
it is straightforward to show (for example~\cite{NimaEFT}) eq.(\ref{Hull}) implies that det$[K_{n}]\geq0$ for all $n=4N{+}2$. We will now explore what this positivity implies for the allowed values of $z_i$ for the amplitudes we computed in eq.(\ref{OneLoopAmp}).

Note that there are three scales involved, $M_{pl} \gg \Lambda \gg m_i$, where $\Lambda$ represents the cutoff for which gravity is UV completed.  Unlike the corresponding analysis in four dimensions, where eq.(\ref{3Dpara}) would be parametrized by $z_i$'s and $M_{pl}$ only (see e.g.~\cite{Cheung:2014ega}), in three dimensions the contributions from the physics above $\Lambda$ would enter into eq.(\ref{3Dpara}) by replacing $m_i\rightarrow \Lambda$ and thus are negligible.

The reader might wonder why there is any constraint at all, given that low energy polynomial amplitude \textit{has} a UV completion, the one-loop massive amplitude. The non-triviality comes in that the one-loop massive amplitude behaves as $ s^2$ in the high energy due to the gravitational effects, as shown previously. Thus the fact that eq.(\ref{Hull}) involving $g_2$ requires that the Froissart bound to hold, and therefore the condition that gravity is UV completed is incorporated~\cite{Noneh2}.

%%%%%%%%%%%%%%%%%%%%%%%%%%%%%%%%%%%%%%%%%%%%%%
\section{Constraints on light states}  
\begin{figure}
\begin{center}
\includegraphics[scale=0.45]{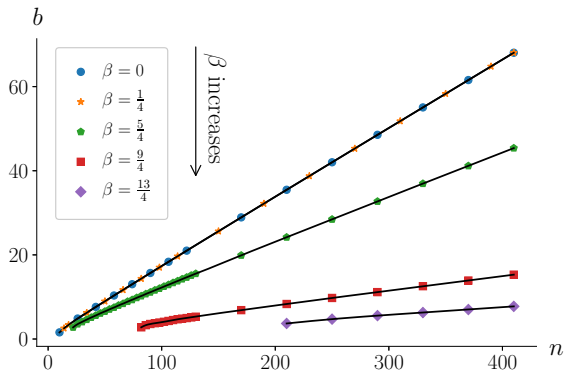}
\caption{ The plot of $b$ with respect to different values of charged to neutral state mass ratio $\beta$, with $\beta=(0,\frac{1}{4},\frac{5}{4},\frac{9}{4},\frac{13}{4})$ from top to bottom. Note that $\beta=0$ corresponds to the case where one has an isolated charged state.}
\label{Plot2}
\end{center}
\end{figure}

\begin{figure}
\begin{center}
\includegraphics[scale=0.45]{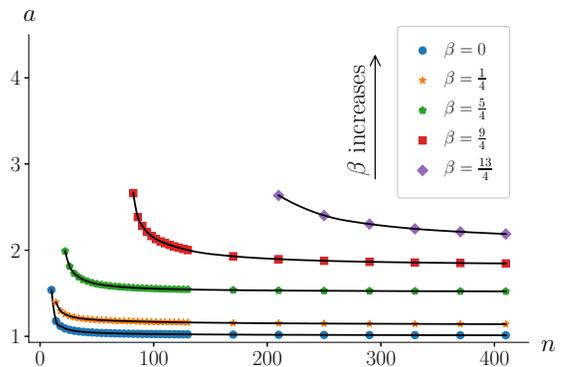}
\caption{ The plot of $a$ with respect to different values of charged to neutral state mass ratio $\beta$, increasing from bottom to top. We see that at sufficiently large $n$ it asymptotes to a fixed value, denoted as $a_{asymp}(\beta)$.}
\label{Plot3}
\end{center}
\end{figure}

%%%%%%%%%%%%%%%%%%%%%%%%%%%%%%%%%%%%%%%%%%%%%%
Let's study the implications of the Hankel matrix constraints, by beginning with the case where the spectrum contains an isolated charged state that is light, therefore it dominates the contribution to $g_n$. In this case $g_n$ takes the following form, 
\eq
 \vspace{-0.2cm} g_n=\frac{c_{n,4}z^4{+}c_{n,2}z^2{+}c_{n,0}}{m^{2n{-}3} M^2_{pl}}\,,
\eqe
where the $c_n$'s are given in eq.(\ref{ScalarCoeff}) and eq.(\ref{FermiCoeff}) depending on whether the light state is a fermion or a scalar. Here, we will present the analysis for an isolated fermion. The corresponding results for a scalar are qualitatively the same. The positivity of det$[K_{n}]$ now becomes a constraint on the $z$ of this light state. For example positivity of det$[K_{206}]$ implies:
\eq
0<|z| < 1.02,\, {\rm or}\;\;  |z| > 34.82\,.
\eqe
In general as $n$ increases, $K_{n}\rightarrow K_{n+4}$, one finds new bounds that are a subset of previous ones. Thus the exact bound is given by det$[K_\infty]>0$. There are two regions of viability which we denote as $|z|<a$ and $|z|>b$. As it is difficult at this stage to obtain the bound from det$[K_\infty]>0$, we instead extrapolate how $a,b$ behaves with respect to $n$ as shown in fig.(\ref{Plot2}) and fig.(\ref{Plot3}). We see that $a$ is asymptotically approaching a fixed point somewhere above $1$, while $b$ is simply rising linearly.  Extrapolate to infinity, we arrive at the conclusion that $|z|$ simply cannot be greater than $1$~\cite{large-n}. 

We see that a spectrum containing a light state with $|z|>1$ is inconsistent with unitarity. This appears to be the opposite of the weak gravity conjecture, where any spectrum with $|z|>1$  satisfies the conjecture. Again, this does not necessarily indicate any inconsistency associated with the compactified Standard Model, as there the spectrum of light states consists of electrons \textit{and} neutrinos. In the next subsection we will see that the inclusion of other light states does alleviate this tension. Again as stressed previously, this is a reflection of the low energy amplitude being parameterized by both $\{z_i, m_i\}$, as opposed to just $z_i$ in the analysis for the four-dimensional theories~\cite{Cheung:2014ega}.  

We have also performed analogue analysis for the cases with multiple-$U(1)$ gauge theories coupled with gravity by considering the low-energy expansion on the crossing symmetric amplitudes as defined in eq.(\ref{eq:multi-U(1)}), and similar results have been found, see the discussion at \cite{MultiU(1)}.  
%%%%%%%%%%%%%%%%%%%%%%%%%%%%%%%%%%%%%%%%%%%%%%
\subsection{The addition of light state  $\frac{m_e}{m_0}>1$}  
%%%%%%%%%%%%%%%%%%%%%%%%%%%%%%%%%%%%%%%%%%%%%%

We now consider the inclusion of a light neutral state along with a charged state. We introduce $\beta\equiv \frac{m_e}{m_0}$, parameterizing the mass ratio of the charged $m_e$ to neutral state $m_0$. In this case $g_n$ takes the form:
\eq
 g_n=\frac{c_{n,4}z^4{+}c_{n,2}z^2{+}(1+\beta^{2n{-}3})c_{n,0}}{m_e^{2n{-}3} M^2_{pl}}\,.
\eqe   
Plotting $(a, b)$ against the number of derivatives with respect to a fixed $\beta$, we find that  for $\beta\leq1$ the plot for $a,b$ is near identical to that without the neutral state as one can compare with the orange plots ($\beta=\frac{1}{4}$) with the blue plots ($\beta=0$) in fig.(\ref{Plot2}) and fig.(\ref{Plot3}). For $\beta>1$ the upper bound $a$ increases as one raises $\beta$, while the slope for the linear rising of $b$ decreases. This shows that the tension for isolated $|z|>1$ states and unitarity is alleviated with the inclusion of a light neutral state with $\beta>1$~\cite{MultiU(1)-2}.

Given that the constraints always come in the pair $(a,b)$, if we assume that the observed trend of linear rise with respect to $n$ persists to $n\rightarrow\infty$, we would conclude that the allowed region for $z$ will be solely determined by the asymptotic value of $a$ for a fixed $\beta$, denoted as $a_{asymp}(\beta)$:
\eq
\framebox[5cm][c]{$ 0<|z|\leq a_{asymp}(\beta)$\,.}
\eqe 
From the plots in fig.(\ref{Plot3}), we see that at $n\sim400$, the value of $a$ is stabilized, and gives a good approximation to $a_{asymp}(\beta)$. We find:
\eqa
a_{asymp}\left(0\right)= 1.01\,,
\quad
a_{asymp}\left(\tfrac{1}{4}\right)=       1.14\,,    \nonumber\\  
a_{asymp}\left(\tfrac{5}{4}\right)=       1.52\,, \quad
a_{asymp}\left(\tfrac{9}{4}\right)=       1.84\,. 
\eqae
%
%
%\eqa
%a_{asymp}\left(0\right)~&=& 1.012547390204077243838\nonumber\,,\\
%a_{asymp}\left(\frac{1}{4}\right)&=&       1.141244407316497755441\,,    \nonumber\\  
%a_{asymp}\left(\frac{5}{4}\right)&=&       1.521513785699259470499\,, \\
%a_{asymp}\left(\frac{9}{4}\right)&=&       1.845769623559088479472\,. \nonumber
%\eqae
%

We can also consider the case where the additional light state is charged. In this case the low energy couplings $g_n$ are parameterized by:
\eq
\frac{c_{n,4}(z^4{+}\beta^{2n{-}3}z'^4){+}c_{n,2}(z^2{+}\beta^{2n{-}3}z'^2){+}(1+\beta^{2n{-}3})c_{n,0}}{m_e^{2n{-}3} M^2_{pl}}\,,
\eqe 
where $z'$ is the charge-to-mass ratio of the additional state. It is straightforward to check that for fixed $\beta$, as one increases in $z'$, $(a,\frac{1}{b})$ decrease with the maximum given by $z'=0$, i.e., the neutral state.

\vspace{-0.2cm}

%%%%%%%%%%%%%%%%%%%%%%%%%%%%%%%%%%%%%%%%%      
\section{Outlook}
%%%%%%%%%%%%%%%%%%%%%%%%%%%%%%%%%%%%%%%%%
By studying the infinite set of positivity constraints on the Hankel matrices, we find that an isolated light charged state with $|z|>1$ is inconsistent with unitarity, unless there exists another state whose mass is lighter than the one with $|z|>1$. If the state is neutral, then there is an upper bound on the charged to neutral mass ratio $\beta$, given by the solution to 
\eq
z=a_{symp}(\beta)\,.
\eqe 
This provides an interesting connection between the pattern of light states in the spectrum and UV completion of quantum gravity. A better understanding of the function $a_{symp}(\beta)$ in the future can provide an theoretical bound on the neutrino mass in the three-dimensional analogue of the Standard Model, by inserting the electron $|z_e|\sim 10^{22}$ in the LHS of the above equation.

\vspace{-0.4cm}

%%%%%%%%%%%%%%%%%%%%%%%%%%%%%%%%%%%%%%%%%%%%%%%%%%%%%
\section{Acknowledgements}
%%%%%%%%%%%%%%%%%%%%%%%%%%%%%%%%%%%%%%%%%%%%%%%%%%%%
We thank Clifford Cheung for the collaboration at the early stage of this project. We are also grateful to Nima Arkani-Hamed, Daniel Junghans, Grant N. Remmen and Gary Shiu for helpful discussions and comments on the early version of the draft. W.-M. Chen is supported by MoST Grant No. 107-2811-M-007-028. Y-t Huang is supported by  MoST Grant No. 106-2628-M-002-012-MY3. T.N. is supported in part by JSPS KAKENHI Grant Numbers JP17H02894 and JP18K13539, and MEXT KAKENHI Grant Number JP18H04352. C. Wen is supported by a Royal Society University Research Fellowship No. UF160350. This research was supported in part by the National Science Foundation under Grant No. NSF PHY17-48958.

\vspace{-0.3cm}
\section*{Appendix: Extracting integral coefficients in three dimensions}
In three dimensions, a loop momentum can be fixed completely by imposing three on-shell conditions,
which reflects that an one-loop integral can be represented on the basis of triangle, bubble and tadpole scalar integrals. Here we present a method to extract integral coefficients for scattering amplitudes in three dimensions. It resembles to that was developed in four dimensions \cite{Forde:2007mi}. 

\begin{figure}[t]
\begin{minipage}{0.2\textwidth}
\begin{tabular}{c}
\includegraphics[scale=0.4]{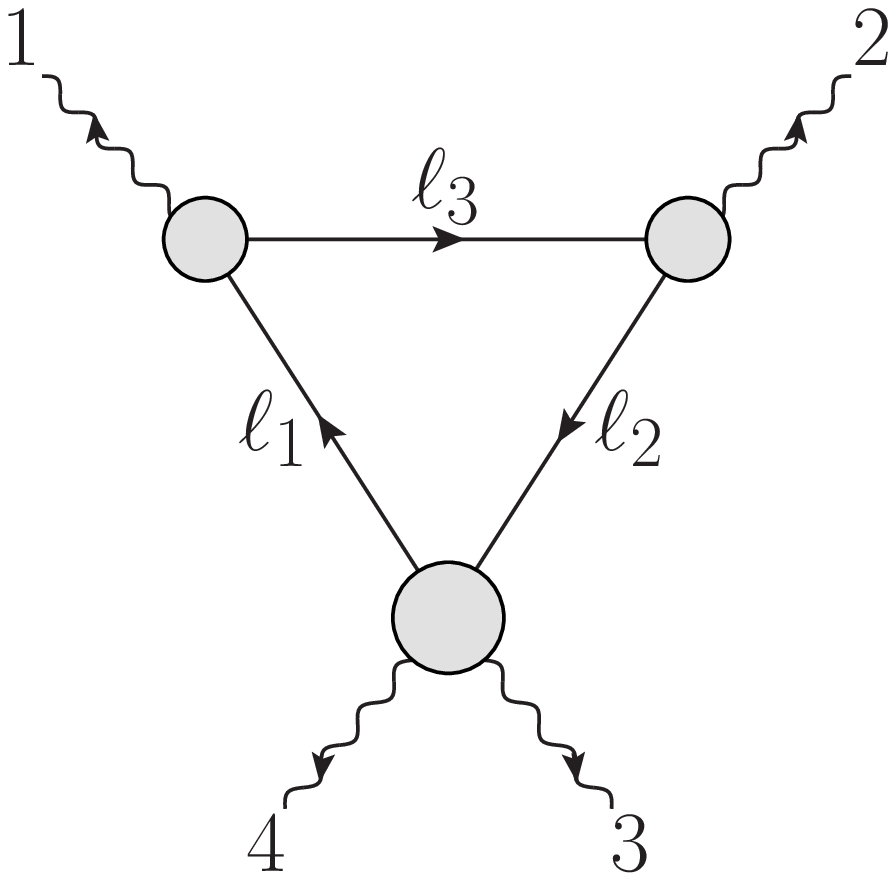}\\
(a)
\end{tabular}
\end{minipage}~~~~~~~~~~\begin{minipage}{0.2\textwidth}
\begin{tabular}{c}
\includegraphics[scale=0.4]{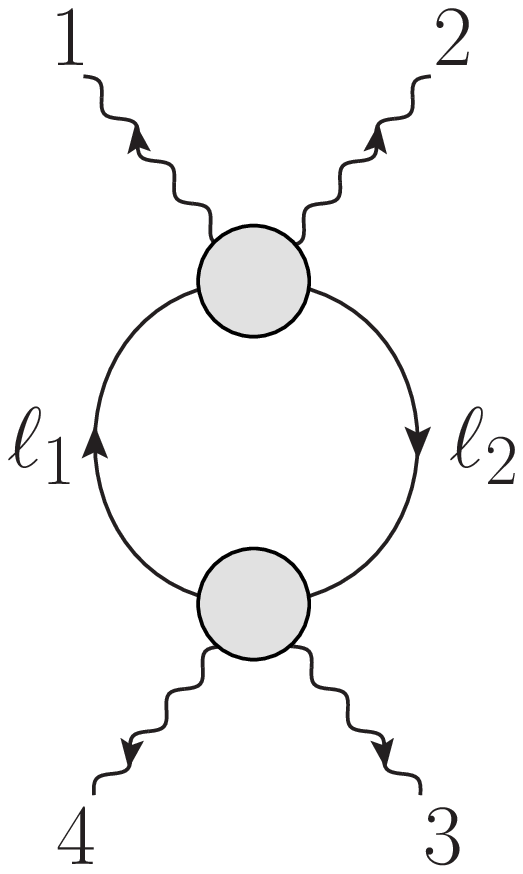}\\
(b)
\end{tabular}
\end{minipage}
\caption{the cut diagrams for the extraction of triangle and bubble coefficients: (a) triple-cut (b) double-cut. All exposed lines are put on-shell.}\label{ccut}
\end{figure}

We are interested in the four-photon one-loop amplitudes in Einstein-Maxwell theory with massive fermionic or scalar matters in the loop. The tadpole will be ignored in our discussion because its contribution is proportional to $m$ (simply by power counting), therefore it is irrelevant to the discussion of large-mass expansion in the main text of the article. Moreover, ``massless bubble", the bubble diagram with single massless leg on one side, is also ignored. ``Massless bubble" gives a non-local contribution, but that just means the non-local pieces from the large mass expansion of triangles and bubbles need to be removed, such that the principle of EFT is at work. Therefore for our purpose, we will focus on bubble and triangle coefficients which can be extracted by considering loop integrals on cuts, which can be done by Feynman diagrams with cut conditions imposed or taking a product of on-shell tree amplitudes, as shown in fig.(\ref{ccut}). 

Adapt the convention in \cite{Brandhuber:2012un}, the four external momenta are represented as $p_i^{\alpha\beta}=\lambda_i^\alpha \lambda_i^\beta$ with $i=1,2,3,4$ and the loop momentum $\ell_1$ is parametrized by two parameters $x_1$ and $x_2$ as
\beqa
\hspace*{-0.4cm}\label{loopp}\ell_1^{\alpha\beta}= x_1 \lambda_1^{\alpha}\lambda_1^{\beta}+(1-x_1)\lambda_2^{\alpha}\la_2^{\beta}+x_2 (\lambda_1^{\alpha}\lambda_2^{\beta}+\lambda_2^{\alpha}\lambda_1^{\beta})\,,
\eeqa
where $\alpha$ is the $\mathrm{SL(2)}$ spinor index. The relevant four-point amplitude for two-photons and two massive scalars are given as:
\eqa
M_{qed}^{\rm tree}(1^\gamma 2^\gamma 3^\phi 4^\phi)&=&\frac{z^2 m^2}{M_{pl}}\left[\frac{s\, m^2}{(t-m^2)(u-m^2)}{-}\frac{1}{2}\right]\nonumber\,,\\
M_{grav}^{\rm tree}(1^\gamma 2^\gamma 3^\phi 4^\phi)&=&\frac{1}{4M_{pl}}\frac{(t-m^2)(u-m^2)}{s}\,,
\eqae
where the subscript indicates the first is the QED process while the second is due to graviton exchange. It is instructive to see that the graviton exchange,
$$\includegraphics[scale=0.4]{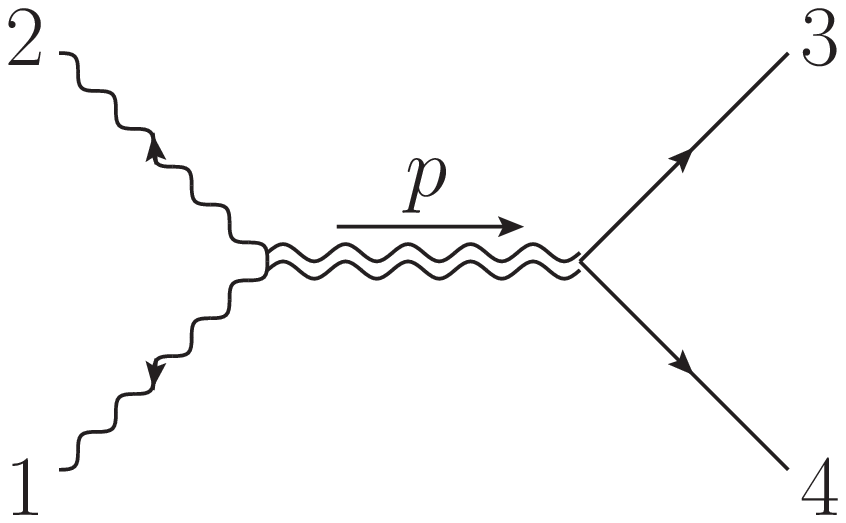}\,,$$
does not have a residue. For $s=0$, $p_1^2=p_2^2=p_1\cdot p_2=0$ and hence $p_1,p_2$ are collinear in three-dimensions even for complex momenta. On the other side $(p-p_3)^2=p_4^2$, where $p=-p_1-p_2$, leads to $p\cdot p_3=p\cdot p_4=0$ at $s=0$. Thus in the $s$-channel on-shell kinematics:
\eq
(t-m^2)|_{s{=}0}=2(p_2\cdot p_3)|_{s{=}0}\sim \;p\cdot p_3=0\,.
\eqe
Thus one concludes that the $s$-channel residue for the graviton exchange vanishes. On other hand, the amplitude does have a singularity in the zero-momentum graviton exchange, namely when $p_1 + p_2 =0$. Therefore, the amplitude cannot be expressed as contact terms.  

The triangle coefficient $\mathcal C(34;1,2)$ can be obtained by taking the triple cut (as in fig.\ref{ccut}(a)). First of all, the triple-cut condition requires
$(x_1,x_2)=(x_1^s, x_2^\pm)$, where $x_1^s=1$ and $x_2^\pm =\pm \frac{i m}{\braket{1}{2}}$. Substitute  the solutions into the cut, it gives the triangle coefficient,
\beqa
\label{trc}\mathcal C(34;1,2)=\frac{1}{2}\sum_{x_1=x_s\atop{x_2=x_2^\pm}} \Bigg[\mathcal I(x_1,x_2)\Big|_{\mathrm{triple-cut}}\Bigg]\,.
\eeqa
Similarly, we extract the bubble coefficient by considering the double cut (as in fig.\ref{ccut}(b)).
The solutions to double-cut conditions are
\beqa
x_1=x_1^\pm\,,~\mathrm{with}~~x_1^\pm=\frac{1}{2}\pm \sqrt{\frac{1}{4}-\frac{ m^2}{s}-x_2^2}\,.
\eeqa
Use the solutions, the integrand on the double cut can be expressed as  a function of $x_2$, which contains poles at $x_2=x_2^\pm$ relevant to the triangle coefficient. In order to avoid the poles, we expand the double-cut result at $x_1=x_1^\pm$ around $x_2=\infty$,
\beqa
\label{bbc}\mathcal C(12;34)&=&\frac{1}{2}
\sum_{x_1=x_1^\pm} \Bigg[\mathcal I(x_1,x_2)\Big|_{\mathrm{double-cut}}\Bigg]\Bigg|_{{\mathrm{series}\atop{\mathrm{expansion}}}\atop{\mathrm{at}~x_2=\infty}} 
\nonumber \\
&=&\sum_{i=0}^k b_i \int d x_2 J_{x_2}  x_2^i\,,
\eeqa
where $b_i$'s are some coefficients from the expansion and $J_{x_2}$ is the Jacobian from the change of variables. To perform the integration of $x_2$, we will use a procedure which is similar to that in~\cite{Forde:2007mi}. First we utilize the results from Veltman-Passarino reduction
\beqa
\label{fm1}
&&\hspace*{-0.5cm } \int  \frac{d^3 \ell_1}{(2\pi)^3}\frac{[\lambda_1^a (\ell_1)_a^{~b}\lambda_{2, b}]^n}{[\ell_1^2-m^2][ (\ell_1-p_1-p_2)^2-m^2]} 
 \\[0.2cm]
\notag &&\hspace*{-0.4cm}=2^{n-1} (s^2-4 s m^2 )^{\frac{n}{2}}\, (n-1)!!\, {n!!}\,[1+(-1)^n]\, I_{\rm bub}(s, m)\,,
\eeqa
where $I_{\rm bub}(s, m)$ is the scalar bubble integral 
\beqa  \label{eq:bubble-integral}
I_{\rm bub}(s, m)=\int \frac{d^3 \ell_1}{(2\pi)^3}\frac{1}{(\ell_1^2-m^2)(\ell_2^2-m^2)}\,.
\eeqa

Let's now impose the double-cut conditions $\ell_1^2=\ell_2^2=0$ on both sides of eq.\eqref{fm1}. 
The double-cut of the L.H.S. can be obtained by substituting the loop momentum eq.\eqref{loopp} into the numerator of the integrand, which equals to the cut results on the R.H.S.. From that, we find the $x_2$ integration gives the following results
\beqa \label{fm2}
\hspace*{-2mm}\int \! d x_2 J_{x_2}x_2^n =\frac{(1-4  m^2/s )^{n/  2}(n-1)!!}{2^{n+1} n!!}[1+(-1)^n]\,.
\eeqa
Use the results of eqs.\eqref{trc}, \eqref{bbc} and \eqref{fm2}, we can now obtain the triangle and bubble coefficients for the four-photon amplitudes at one loop. Begin with the case of a massive scalar in the loop, we find 
\beqa
&&\hspace*{-1cm}\mathcal C(34;1,2)=-\frac{z^2 m^4 t u \left(4 m^2+s\right)}{4   s^2 M_{pl}^2}
 \\
&& -\frac{z^4 m^6 \left(4 m^2-s\right) \left[2 m^2 \left(t^2+u^2\right)+s t u\right]}{2   \left(4 m^2 u+s t\right) \left(4 m^2 t+s u\right)M_{pl}^2} \,, \nonumber \\
&&\hspace*{-1cm} \mathcal C(12;34)= \frac{8 z^4 s^2 m^4 {+} z^2 m^2[4 m^2  (3 s^2 {-} 32 t u ) {-} s^3]}{32   s^2 M_{pl}^2}  \nonumber  \\
&&+\frac{ (16 m^4 {+} s^2 )  (3 s^2 {-} 8 t u )-8 m^2  (t^3 {+} u^3 {-}11 s t u )}{2048   s^2 M_{pl}^2} \,.  \nonumber
\eeqa
Similarly, for the one-loop amplitude with an internal massive fermion, we have
\beqa
&& \hspace*{-0.5cm}\mathcal C(34;1,2)=\frac{z^2  m^4 \left(16 m^2 t u+3 s t u+t^3+u^3\right)}{16  s^2 M_{pl}^2}  \\
&&+ \frac{ z^4 m^6 \left[32 m^4 (t^2 {+} u^2 )+8 m^2 (t^3 {+}u^3 )-s^2 (s^2 {+} 2 t u )\right]}{8 \left(4 m^2 u+s t\right) \left(4 m^2 t+s u\right)M_{pl}^2}\,, \nonumber \\
&& \hspace*{-0.5cm}\mathcal C(12;34)=\frac{  (4 m^2 {-} s ) \left[4 m^2  (8 t u{-}3 s^2 ){+}5 s t u{+}t^3{+}u^3\right]}{2048   s^2 M_{pl}^2} \nonumber \\
&&+ \frac{8z^4 s^2 m^4 {+} z^2 m^2 \left[4 m^2 (32 t u{-}3 s^2 ){+}5 s t u{+}t^3{+}u^3\right]}{32   s^2 M_{pl}^2}\,. \nonumber
\eeqa
Finally, the complete result of the loop amplitudes is given by 
\beqa
\!\!\!\!\!\! M_4(s,t)& =&4\, \mathcal{C}(34; 1, 2)   I_{\rm tri}(s,m^2)   
+ \mathcal{C}(12;34) I_{\rm bub}(s,m^2)  \nonumber
\\ && ~~~ + (s \leftrightarrow t)+ (s \leftrightarrow u) \,,
\eeqa
where the bubble scalar integral $I_{\mathrm{bub}}(s^2,m)$ is defined in eq.(\ref{eq:bubble-integral}), and the triangle integral is given by 
\beqa
\hspace*{-0.4cm} I_{\mathrm{tri}}(s^2,m)\hspace*{-0.05cm}   = \hspace*{-0.15cm} \int \frac{d^3 \ell_1}{(2\pi)^3}\frac{1}{(\ell_1^2-m^2)(\ell_2^2-m^2)(\ell_3^2-m^2)}\, .
\eeqa

%{99}


\begin{thebibliography}{99}

\bibitem{Adams:2006sv} 
  A.~Adams, N.~Arkani-Hamed, S.~Dubovsky, A.~Nicolis and R.~Rattazzi,
  ``Causality, analyticity and an IR obstruction to UV completion,''
  JHEP {\bf 0610}, 014 (2006)
 % doi:10.1088/1126-6708/2006/10/014
  [hep-th/0602178].
  %%CITATION = doi:10.1088/1126-6708/2006/10/014;%%
  %477 citations counted in INSPIRE as of 30 Dec 2018

\bibitem{NimaEFT}
N.~Arkani-Hamed, T-z~Huang, Y-t~Huang, in preparation.    

\bibitem{ArkaniHamed:2006dz} 
  N.~Arkani-Hamed, L.~Motl, A.~Nicolis and C.~Vafa,
  ``The String landscape, black holes and gravity as the weakest force,''
  JHEP {\bf 0706}, 060 (2007)
 % doi:10.1088/1126-6708/2007/06/060
  [hep-th/0601001].
  %%CITATION = doi:10.1088/1126-6708/2007/06/060;%%
  %462 citations counted in INSPIRE as of 21 Jan 2019  
  
%\cite{Kats:2006xp}
\bibitem{Kats:2006xp} 
  Y.~Kats, L.~Motl and M.~Padi,
  ``Higher-order corrections to mass-charge relation of extremal black holes,''
  JHEP {\bf 0712}, 068 (2007)
 % doi:10.1088/1126-6708/2007/12/068
  [hep-th/0606100].
  %%CITATION = doi:10.1088/1126-6708/2007/12/068;%%
  %78 citations counted in INSPIRE as of 30 Jan 2019

%\cite{Cremonini:2009ih}
\bibitem{Cremonini:2009ih} 
  S.~Cremonini, J.~T.~Liu and P.~Szepietowski,
  ``Higher Derivative Corrections to R-charged Black Holes: Boundary Counterterms and the Mass-Charge Relation,''
  JHEP {\bf 1003}, 042 (2010)
%  doi:10.1007/JHEP03(2010)042
  [arXiv:0910.5159 [hep-th]].
  %%CITATION = doi:10.1007/JHEP03(2010)042;%%
  %23 citations counted in INSPIRE as of 30 Jan 2019

%\cite{Cheung:2014ega}
\bibitem{Cheung:2014ega} 
  C.~Cheung and G.~N.~Remmen,
  ``Infrared Consistency and the Weak Gravity Conjecture,''
  JHEP {\bf 1412}, 087 (2014)
 % doi:10.1007/JHEP12(2014)087
  [arXiv:1407.7865 [hep-th]].
  %%CITATION = doi:10.1007/JHEP12(2014)087;%%
  %29 citations counted in INSPIRE as of 25 Jul 2018

%\cite{Cheung:2018cwt}
\bibitem{Cheung:2018cwt} 
  C.~Cheung, J.~Liu and G.~N.~Remmen,
  ``Proof of the Weak Gravity Conjecture from Black Hole Entropy,''
  JHEP {\bf 1810}, 004 (2018)
 % doi:10.1007/JHEP10(2018)004
  [arXiv:1801.08546 [hep-th]].
  %%CITATION = doi:10.1007/JHEP10(2018)004;%%
  %17 citations counted in INSPIRE as of 30 Jan 2019

%\cite{Andriolo:2018lvp}
\bibitem{Andriolo:2018lvp} 
  S.~Andriolo, D.~Junghans, T.~Noumi and G.~Shiu,
  ``A Tower Weak Gravity Conjecture from Infrared Consistency,''
  Fortsch.\ Phys.\  {\bf 66}, no. 5, 1800020 (2018)
 % doi:10.1002/prop.201800020
  [arXiv:1802.04287 [hep-th]].
  %%CITATION = doi:10.1002/prop.201800020;%%
  %5 citations counted in INSPIRE as of 25 Jul 2018

%\cite{Hamada:2018dde}
\bibitem{Hamada:2018dde} 
  Y.~Hamada, T.~Noumi and G.~Shiu,
  ``Weak Gravity Conjecture from Unitarity and Causality,''
  arXiv:1810.03637 [hep-th].
  %%CITATION = ARXIV:1810.03637;%%
  %6 citations counted in INSPIRE as of 22 Jan 2019

  
\bibitem{commentzstar}
{In~\cite{Cheung:2014ega} the bound $z_\ast$ was obtained by 
absorbing the contribution of  $c_{2,0}$ (which is defined in eq.(\ref{3Dpara})) into the contributions from UV states, denoted as $\gamma$ in~\cite{Cheung:2014ega}, and further assuming $\gamma$ is small. }
\bibitem{Bern:1994zx} 
  Z.~Bern, L.~J.~Dixon, D.~C.~Dunbar and D.~A.~Kosower,
  ``One loop n point gauge theory amplitudes, unitarity and collinear limits,''
  Nucl.\ Phys.\ B {\bf 425}, 217 (1994)
%  doi:10.1016/0550-3213(94)90179-1
  [hep-ph/9403226].
  %%CITATION = doi:10.1016/0550-3213(94)90179-1;%%
  %1093 citations counted in INSPIRE as of 03 Jan 2019
    Z.~Bern, L.~J.~Dixon, D.~C.~Dunbar and D.~A.~Kosower,
  ``Fusing gauge theory tree amplitudes into loop amplitudes,''
  Nucl.\ Phys.\ B {\bf 435}, 59 (1995)
%  doi:10.1016/0550-3213(94)00488-Z
  [hep-ph/9409265].
  %%CITATION = doi:10.1016/0550-3213(94)00488-Z;%%
  %814 citations counted in INSPIRE as of 03 Jan 2019  
\bibitem{compactify} 
One could expect that such a three-dimensional theory is obtained by compactifying the Standard Model in our real world. However, for its justification, we would need some more assumptions on UV completion. It is beyond our scope in the present paper, leaving it for future work. Also note that we assume that the 3D electron has the same charge-to-mass ratio as the Standard Model one because it is invariant under the dimensional reduction.

\bibitem{Cheung:2014vva} 
  C.~Cheung and G.~N.~Remmen,
  ``Naturalness and the Weak Gravity Conjecture,''
  Phys.\ Rev.\ Lett.\  {\bf 113}, 051601 (2014)
%  doi:10.1103/PhysRevLett.113.051601
  [arXiv:1402.2287 [hep-ph]].
  %%CITATION = doi:10.1103/PhysRevLett.113.051601;%%
  %87 citations counted in INSPIRE as of 13 Jan 2019  
  
\bibitem{forward-limit}
{We use the notation $s=(p_1+p_2)^2$, and $t=(p_1{+}p_4)^2=\frac{-s(1-\cos\theta)}{2}$, where $\theta$ is the scattering angle in the center of mass frame.}
  
\bibitem{Chen:2015hpa} 
  W.~M.~Chen, Y.~t.~Huang and C.~Wen,
  ``Exact coefficients for higher dimensional operators with sixteen supersymmetries,''
  JHEP {\bf 1509}, 098 (2015)
 % doi:10.1007/JHEP09(2015)098
  [arXiv:1505.07093 [hep-th]].
  %%CITATION = doi:10.1007/JHEP09(2015)098;%%
  %14 citations counted in INSPIRE as of 30 Dec 2018
  
  
  
\bibitem{Ritz:1995nt} 
  A.~Ritz and R.~Delbourgo,
  ``The Low-energy effective Lagrangian for photon interactions in any dimension,''
  Int.\ J.\ Mod.\ Phys.\ A {\bf 11}, 253 (1996)
 % doi:10.1142/S0217751X96000122
  [hep-th/9503160].
  %%CITATION = doi:10.1142/S0217751X96000122;%%
  %13 citations counted in INSPIRE as of 21 Jan 2019  
\bibitem{Drummond:1979pp} 
  I.~T.~Drummond and S.~J.~Hathrell,
  ``QED Vacuum Polarization in a Background Gravitational Field and Its Effect on the Velocity of Photons,''
  Phys.\ Rev.\ D {\bf 22}, 343 (1980).
 % doi:10.1103/PhysRevD.22.343
  %%CITATION = doi:10.1103/PhysRevD.22.343;%%
  %385 citations counted in INSPIRE as of 21 Jan 2019    
 \bibitem{Froissart:1961ux} 
  M.~Froissart,
  ``Asymptotic behavior and subtractions in the Mandelstam representation,''
  Phys.\ Rev.\  {\bf 123}, 1053 (1961).
 % doi:10.1103/PhysRev.123.1053
  %%CITATION = doi:10.1103/PhysRev.123.1053;%%
  %1094 citations counted in INSPIRE as of 30 Dec 2018  
 
 \bibitem{tpole}
The massless $t$-channel pole in the forward limit takes the form~$\frac{s}{\theta}$. We can simply subtract this singularity from the amplitude without affecting the low-energy $\mathcal{O}(s^2)$ coefficient. Also we assume that the subtracted forward amplitude is bounded as $<s^2$ in high energy. Obviously, string theory satisfies this criterion. Besides, it is the case at least for physical momentum configurations, e.g., when the total amplitude satisfies the unitarity bound or when the theory is asymptotic free.

\bibitem{Noneh2} Indeed if one were to study the Hankel matrix that does not involve $g_2$, there are no none-trivial constraints on $z$.   


    \bibitem{large-n}
 One might worry that for a sufficient large $n$, the one-loop coefficients $g_n$ become comparable or even smaller than the leading contributions from the UV states. This is indeed true, however due to the fact that the constraints are in the form of determinants, one can see that UV contributions to the Hankel matrix constraints remain suppressed even for large $n$. 
      
    \bibitem{MultiU(1)} 
In the cases of multiple $U(1)$'s, we introduce auxiliary unit vectors for the photons amplitudes as done in (\ref{eq:multi-U(1)}). We find that the unitarity constraints imply bounds on the charge-to-mass ratio of the form analogue to that of the single-$U(1)$ case in eq.(\ref{eq:ab-bounds}): 
\begin{align*}
0 < |\vec{z}\cdot u|^2 + |\vec{z}\cdot v|^2 < a, \quad   b< |\vec{z}\cdot u|^2 + |\vec{z}\cdot v|^2 ,
\end{align*}
where the upper boundary $a$ tends to $2$ while the lower boundary $b$ approaches to infinity when we impose Hankel matrix constraints as the number of derivatives increases. 

\bibitem{MultiU(1)-2} 
{
A similar conclusion can be obtained also from positivity in multiple-$U(1)$ cases. As studied in~\cite{Andriolo:2018lvp}, positivity in scattering of two different $U(1)$ gauge bosons leads to an upper bound $z^2<2$, which excludes charged particles with $|z|>\sqrt{2}$. Ref.~\cite{Andriolo:2018lvp} then argued that this situation may be reconciled by adding bifundamental particles charged under both $U(1)$'s. Actually, just in our present case, it turns out that sufficiently light neutral particles provide another way out to recover positivity. Note that bifundamental particles do not help in our case, so that our constraints are more stringent.}


  
\bibitem{Forde:2007mi} 
  D.~Forde,
  ``Direct extraction of one-loop integral coefficients,''
  Phys.\ Rev.\ D {\bf 75}, 125019 (2007)
 % doi:10.1103/PhysRevD.75.125019
  [arXiv:0704.1835 [hep-ph]].
  %%CITATION = doi:10.1103/PhysRevD.75.125019;%%
  %283 citations counted in INSPIRE as of 23 Jan 2019  
  %\cite{Brandhuber:2012un}
\bibitem{Brandhuber:2012un} 
  A.~Brandhuber, G.~Travaglini and C.~Wen,
  ``A note on amplitudes in N=6 superconformal Chern-Simons theory,''
  JHEP {\bf 1207}, 160 (2012)
 % doi:10.1007/JHEP07(2012)160
  [arXiv:1205.6705 [hep-th]].
  %%CITATION = doi:10.1007/JHEP07(2012)160;%%
  %25 citations counted in INSPIRE as of 23 Jan 2019

\end{thebibliography}
\end{document}